\@twosidetrue
\@twocolumntrue
\@mparswitchtrue
\def\ds@draft{\overfullrule 5pt}
\def\ds@twocolumn{\@twocolumntrue}
\def\ds@onecolumn{\@twocolumnfalse}
\newif\ifSFB@landscape
\def\ds@landscape{\SFB@landscapetrue}
\newif\ifSFB@galley
\def\ds@galley{\SFB@galleytrue}
\newif\ifSFB@referee
\def\ds@referee{%
 \SFB@refereetrue
 \@twocolumnfalse
}
\@options
\ifSFB@referee
 
\else
 
\fi
\if@twocolumn
  \def\@normalsize{\@setsize\normalsize{11pt}\ixpt\@ixpt
   \abovedisplayskip 6pt plus 2pt minus 2pt
   \belowdisplayskip \abovedisplayskip
   \abovedisplayshortskip 6pt plus 2pt
   \belowdisplayshortskip \abovedisplayshortskip
   \let\@listi\@listI}
 \else
 \ifSFB@referee
  \def\@normalsize{\@setsize\normalsize{14pt}\xiipt\@xiipt
   \abovedisplayskip 4pt plus 1pt minus 1pt
   \belowdisplayskip \abovedisplayskip
   \abovedisplayshortskip 4pt plus 1pt
   \belowdisplayshortskip \abovedisplayshortskip
   \let\@listi\@listI}
 \else
  \def\@normalsize{\@setsize\normalsize{12pt}\ixpt\@ixpt
   \abovedisplayskip 4pt plus 1pt minus 1pt
   \belowdisplayskip \abovedisplayskip
   \abovedisplayshortskip 4pt plus 1pt
   \belowdisplayshortskip \abovedisplayshortskip
   \let\@listi\@listI}
 \fi
\fi
\def\small{\@setsize\small{10pt}\viiipt\@viiipt
 \abovedisplayskip 4pt plus 1pt minus 1pt
 \belowdisplayskip \abovedisplayskip
 \abovedisplayshortskip 4pt plus 1pt
 \belowdisplayshortskip \abovedisplayshortskip
 \def\@listi{\leftmargin\leftmargini
  \topsep 2pt plus 1pt minus 1pt
  \parsep \z@
  \itemsep 2pt}}
\def\footnotesize{\@setsize\footnotesize{10pt}\viiipt\@viiipt
 \abovedisplayskip 4pt plus 1pt minus 1pt
 \belowdisplayskip \abovedisplayskip
 \abovedisplayshortskip 4pt plus 1pt
 \belowdisplayshortskip \abovedisplayshortskip
 \def\@listi{\leftmargin\leftmargini
  \topsep 2pt plus 1pt minus 1pt
  \parsep \z@
  \itemsep 2pt}}
\def\scriptsize{\@setsize\scriptsize{8pt}\viipt\@viipt}
\def\tiny{\@setsize\tiny{6pt}\vpt\@vpt}
\if@twocolumn
  \def\large{\@setsize\large{11pt}\xpt\@xpt}
 \else
  \def\large{\@setsize\large{12pt}\xpt\@xpt}
 \fi
\def\Large{\@setsize\Large{14pt}\xiipt\@xiipt}
\def\LARGE{\@setsize\LARGE{17pt}\xivpt\@xivpt}
\def\huge{\@setsize\huge{20pt}\xviipt\@xviipt}
\def\Huge{\@setsize\huge{25pt}\xxpt\@xxpt}
\normalsize

\if@twocolumn
\else
 \ifSFB@referee
  \oddsidemargin \z@  \evensidemargin \z@
 \else
 \fi
\fi
 \topmargin \z@
\newdimen\SFB@measure
\textwidth \SFB@measure
\ifSFB@landscape
 \textwidth \textheight
 \textheight \SFB@measure
\fi
\ifSFB@referee
\fi
\skip\footins 19.5pt plus 12pt minus 1pt

\parskip \z@ plus .1pt
\@beginparpenalty -\@lowpenalty
\@endparpenalty -\@lowpenalty
\@itempenalty -\@lowpenalty
\newcounter{part}
\newcounter {section}
\newcounter {subsection}[section]
\newcounter {subsubsection}[subsection]
\newcounter {paragraph}[subsubsection]
\newcounter {subparagraph}[paragraph]
\def\thepart          {\arabic{part}}
\def\thesection       {\arabic{section}}

\def\part{\par \addvspace{4ex}\@afterindentfalse
 \secdef\@part\@spart}
\def\@part[#1]#2{\ifnum \c@secnumdepth >\m@ne
  \refstepcounter{part}
  \addcontentsline{toc}{part}{Part \thepart: #1}
 \else \addcontentsline{toc}{part}{#1}
 \fi
 {\parindent 0pt \raggedright
  \ifnum \c@secnumdepth >\m@ne
   \large\rm PART
   \ifcase\thepart \or ONE \or TWO \or THREE \or FOUR \or FIVE
    \or SIX \or SEVEN \or EIGHT \or NINE \or TEN \else \fi
   \par \nobreak
  \fi
  \LARGE \rm #2 \markboth{}{}\par }
 \nobreak \vskip 3ex \@afterheading}
\def\@spart#1{{\parindent 0pt \raggedright
  \LARGE \rm #1\par}
 \nobreak \vskip 3ex \@afterheading}

\def\section{\@startsection {section}{1}{\z@}
 {-24pt plus -12pt minus -1pt}
 {6pt}
 {\SFB@hangraggedright\normalsize\bf}}
\def\subsection{\@startsection{subsection}{2}{\z@}
 {-18pt plus -9pt minus -1pt}
 {6pt}
 {\SFB@hangraggedright\large\bf}}
\def\subsubsection{\@startsection{subsubsection}{3}{\z@}
 {-18pt plus -9pt minus -1pt}
 {6pt}
 {\SFB@hangraggedright\normalsize\it}}
\def\paragraph{\@startsection{paragraph}{4}{\z@}
 {12pt plus 2.25pt minus 1pt}{-0.5em}{\normalsize\bf}}
\def\subparagraph{\@startsection{subparagraph}{5}{\parindent}
 {12pt plus 2.25pt minus 1pt}{-0.5em}{\normalsize\it}}
\def\SFB@hangraggedright{\rightskip\@flushglue \let\\=\newline}
\def\@sect#1#2#3#4#5#6[#7]#8{%
 \ifnum #2>\c@secnumdepth
  \def\@svsec{}%
 \else
  \refstepcounter{#1}
  \ifnum #2=\@ne
   \ifSFB@appendix \edef\@svsec{}%
             \else \edef\@svsec{\csname the#1\endcsname\hskip 1em}%
   \fi
  \else \edef\@svsec{\csname the#1\endcsname\hskip 1em}%
  \fi
 \fi
 \@tempskipa #5\relax
 \ifdim \@tempskipa>\z@
  \begingroup #6\relax
   \ifnum #2=\@ne
    \ifSFB@appendix
     \@hangfrom{\hskip #3\relax\@svsec}{\interlinepenalty \@M
      APPENDIX \csname the#1\endcsname:\hskip 0.5em\uppercase{#8}\par}%
    \else
     \@hangfrom{\hskip #3\relax\@svsec}{\interlinepenalty \@M
      \uppercase{#8}\par}%
    \fi
   \else
    \@hangfrom{\hskip #3\relax\@svsec}{\interlinepenalty \@M #8\par}%
   \fi
  \endgroup
  \csname #1mark\endcsname{#7}%
  \addcontentsline{toc}{#1}{\ifnum #2>\c@secnumdepth \else
   \protect\numberline{\csname the#1\endcsname}\fi #7}%
 \else
  \def\@svsechd{#6\hskip #3\@svsec \ifnum #2=\@ne\uppercase{#8}\else #8\fi
  \csname #1mark\endcsname{#7}
  \addcontentsline{toc}{#1}{\ifnum #2>\c@secnumdepth \else
   \protect\numberline{\csname the#1\endcsname}\fi#7}}%
 \fi
 \@xsect{#5}}

\newif\ifSFB@appendix
\def\appendix{\par
 \SFB@appendixtrue
 \setcounter{section}{0}
 \def\thesection{A\arabic{section}}
 \setcounter{equation}{0}
 \def\theequation{A\arabic{equation}}
 \setcounter{figure}{0}
 \def\thefigure{A\@arabic\c@figure}
 \setcounter{table}{0}
 \def\thetable{A\@arabic\c@table}
}

\newskip\@indentskip
\newskip\smallindent
\newskip\@footindent
\newskip\@leftskip
\@footindent=\smallindent
\@leftskip=\z@

\leftmargini   \@indentskip
\leftmargin\leftmargini
\labelwidth\leftmargini\advance\labelwidth-\labelsep
\def\makeRRlabel#1{\hss\llap{#1}}
\def\@listI{\leftmargin\leftmargini
 \parsep \z@
 \topsep 6pt plus 1pt minus 1pt
 \itemsep \z@ plus .1pt
}
\let\@listi\@listI
\@listi
\def\@listii{\leftmargin\leftmarginii
 \labelwidth\leftmarginii\advance\labelwidth-\labelsep
 \topsep 6pt plus 1pt minus 1pt
 \parsep \z@
 \itemsep \z@ plus .1pt
}
\def\@listiii{\leftmargin\leftmarginiii
 \labelwidth\leftmarginiii\advance\labelwidth-\labelsep
 \topsep 6pt plus 1pt minus 1pt
 \parsep \z@
 \partopsep \z@
 \itemsep \topsep
}
\def\@listiv{\leftmargin\leftmarginiv
 \labelwidth\leftmarginiv\advance\labelwidth-\labelsep
}
\def\@listv{\leftmargin\leftmarginv
 \labelwidth\leftmarginv\advance\labelwidth-\labelsep
}
\def\@listvi{\leftmargin\leftmarginvi
 \labelwidth\leftmarginvi\advance\labelwidth-\labelsep
}
\def\itemize{\ifnum \@itemdepth >3 \@toodeep
  \else \advance\@itemdepth \@ne
   \edef\@itemitem{labelitem\romannumeral\the\@itemdepth}%
   \list{\csname\@itemitem\endcsname}%
    {\let\makelabel\makeRRlabel}%
  \fi}

\def\enumerate{\ifnum \@enumdepth >3 \@toodeep \else
  \advance\@enumdepth \@ne
  \edef\@enumctr{enum\romannumeral\the\@enumdepth}%
 \fi
 \@ifnextchar [{\@enumeratetwo}{\@enumerateone}%
}
\def\@enumeratetwo[#1]{%
 \list{\csname label\@enumctr\endcsname}%
  {\settowidth\labelwidth{[#1]}
   \leftmargin\labelwidth \advance\leftmargin\labelsep
   \usecounter{\@enumctr}
   \let\makelabel\makeRRlabel}
}
\def\@enumerateone{%
 \list{\csname label\@enumctr\endcsname}%
  {\usecounter{\@enumctr}
   \let\makelabel\makeRRlabel}}

\def\theenumi{(\roman{enumi})}

\def\theenumii{(\alph{enumii})}
\def\p@enumii{\theenumi}

\def\theenumiii{(\arabic{enumiii})}
\def\p@enumiii{\theenumi(\theenumii)}

\def\p@enumiv{\p@enumiii\theenumiii}
\def\description{\list{}{\labelwidth\z@ \itemindent-\leftmargin
  \leftmargin 1em
  \itemindent-1em
}}

\def\verse{\let\\=\@centercr
 \list{}{\itemsep\z@
  \itemindent -\@indentskip
  \listparindent \itemindent
  \rightmargin\leftmargin
  \advance\leftmargin \@indentskip}\item[]}

\def\quotation{\list{}{\listparindent \smallindent
  \leftmargin\z@\rightmargin\leftmargin
  \parsep 0pt plus 1pt}\item[]\small}

\def\quote{\list{}{\leftmargin\z@\rightmargin\leftmargin}\item[]\small}

\def\@begintheorem#1#2{\rm \trivlist \item[\hskip \labelsep{\bf #1\ #2.}]}
\def\@opargbegintheorem#1#2#3{\rm \trivlist
  \item[\hskip \labelsep{\bf #1\ #2.\ (#3)}]}
\def\@endtheorem{\endtrivlist}
\@namedef{proof*}{\rm \trivlist \item[\hskip \labelsep{\it Proof.}]}
\@namedef{endproof*}{\endtrivlist}

\def\titlepage{\@restonecolfalse\if@twocolumn\@restonecoltrue\onecolumn
  \else \newpage \fi \thispagestyle{empty}\c@page\z@}
\def\endtitlepage{\if@restonecol\twocolumn \else \newpage \fi}

\def\tabular{\def\@halignto{}
 \def\hline{\noalign{\ifnum0=`}\fi
  \vskip 3pt
  \hrule \@height \arrayrulewidth
  \vskip 3pt
  \futurelet \@tempa\@xhline}
 \def\fullhline{\noalign{\ifnum0=`}\fi
  \vskip 3pt
  \hrule \@height \arrayrulewidth
  \vskip 3pt
  \futurelet \@tempa\@xhline}
 \def\@xhline{\ifx\@tempa\hline
   \vskip -6pt
   \vskip \doublerulesep
  \fi
  \ifnum0=`{\fi}}
  \def\@arrayrule{\@addtopreamble{\hskip -.5\arrayrulewidth
                                  \hskip .5\arrayrulewidth}}
\@tabular
}
\tabbingsep \labelsep

\skip\@mpfootins = \skip\footins

\fboxrule = \arrayrulewidth

\def\maketitle{\par
 \begingroup
  \if@twocolumn
   \twocolumn[\vspace*{17pt}\@maketitle]
  \else
   \newpage
   \global\@topnum\z@
   \@maketitle
  \fi
  \thispagestyle{titlepage}
 \endgroup
 \let\maketitle\relax
 \let\@maketitle\relax
 \gdef\@author{}
 \gdef\@title{}
 \let\thanks\relax
}
\def\and{\end{author@tabular}\vskip 6pt\par
 \begin{author@tabular}[t]{@{}l@{}}}
\def\@maketitle{\newpage
 \vspace*{7pt}
 {\raggedright \sloppy
  {\huge \bf \@title \par}
  \vskip 23pt
  {\LARGE
   \begin{author@tabular}[t]{@{}l@{}}\@author
   \end{author@tabular}\par}
  \vskip 22pt
 }
 \par\noindent
 {\small \@date \par}
 \vskip 22pt
}
\def\abstract{\if@twocolumn
  \start@SFBbox\@abstract
 \else
  \@abstract
 \fi}
\def\endabstract{\if@twocolumn
   \endlist\finish@SFBbox
 \else
  \endlist
 \fi}
\def\@abstract{\list{}{\leftmargin 10.5pc\rightmargin\z@
  \parsep 0pt plus 1pt}\item[]\normalsize{\bf ABSTRACT}\\\large} % SFB 0.1.01
\newif\ifSFB@keywords
\def\keywords{\if@twocolumn
  \start@SFBbox\@keywords
 \else
  \@keywords
 \fi
}
\def\@keywords{\list{}{\leftmargin 10.5pc\rightmargin\z@
  \parsep 0pt plus 1pt}\item[]\large{\bf Key words: }}
\def\endkeywords{\if@twocolumn
  \endlist\addvspace{37pt}\finish@SFBbox
 \else
  \endlist
 \fi
 \@thanks
 \gdef\@thanks{}
 \SFB@keywordstrue
}
\def\nokeywords{\ifSFB@keywords\else
 \if@twocolumn \start@SFBbox\addvspace{37pt}\finish@SFBbox \fi
 \@thanks
 \gdef\@thanks{}\fi
}

\def\author@tabular{\def\@halignto{}\@authortable}
\let\endauthor@tabular=\endtabular
\def\author@tabcrone{{\ifnum0=`}\fi\@xtabularcr[-7pt]\small\it
 \let\\=\author@tabcrtwo\ignorespaces}
\def\author@tabcrtwo{{\ifnum0=`}\fi\@xtabularcr[-7pt]\small\it
 \let\\=\author@tabcrtwo\ignorespaces}
\def\@authortable{\leavevmode \hbox \bgroup $\let\@acol\@tabacol
 \let\@classz\@tabclassz \let\@classiv\@tabclassiv
 \let\\=\author@tabcrone \ignorespaces \@tabarray}

\def\start@SFBbox{\@next\@currbox\@freelist{}{}%
 \global\setbox\@currbox
 \vbox\bgroup
  \hsize \textwidth
  \@parboxrestore
}
\def\finish@SFBbox{\par\vskip -\dbltextfloatsep
  \egroup
  \global\count\@currbox\tw@
  \global\@dbltopnum\@ne
  \global\@dbltoproom\maxdimen\@addtodblcol
  \global\vsize\@colht
  \global\@colroom\@colht
}

\mark{{}{}}
\gdef\@author{\mbox{}}
\def\author{\@ifnextchar [{\@authortwo}{\@authorone}}
\def\@authortwo[#1]#2{\gdef\@author{#2}\gdef\@shortauthor{#1}}
\def\@authorone#1{\gdef\@author{#1}\gdef\@shortauthor{#1}}
\gdef\@shortauthor{}
\gdef\@title{\mbox{}}
\def\title{\@ifnextchar [{\@titletwo}{\@titleone}}
\def\@titletwo[#1]#2{\gdef\@title{#2}\gdef\@shorttitle{#1}}
\def\@titleone#1{\gdef\@title{#1}\gdef\@shorttitle{#1}}
\gdef\@shorttitle{}
\def\volume#1{\gdef\@volume{#1}}
\gdef\@volume{000}
\def\microfiche#1{\gdef\@microfiche{#1}}
\gdef\@microfiche{}
\def\pagerange#1{\gdef\@pagerange{#1}}
\gdef\@pagerange{000--000}
\def\journal#1{\gdef\@journal{#1}}
\gdef\@journal{{Mon.\ Not.\ R.\ Astron.\ Soc.} {\bf \@volume}, \@pagerange\
  (\number\year) \@microfiche}
\def\ps@headings{\let\@mkboth\markboth
 \def\@oddhead{\Large \hfill \it \@shorttitle \hspace{1.5em}\rm \thepage}
 \def\@oddfoot{}
 \def\@evenhead{\Large \thepage \hspace{1.5em}\it \@shortauthor \hfill}
 \def\@evenfoot{}
 \def\sectionmark##1{\markboth{##1}{}}
 \def\subsectionmark##1{\markright{##1}}}
\def\ps@myheadings{\let\@mkboth\@gobbletwo
 \def\@oddhead{\Large \it \rightmark \hfill \rm \thepage}
 \def\@oddfoot{}
 \def\@evenhead{\Large \it \leftmark \hfill \rm \thepage}
 \def\@evenfoot{}
 \def\sectionmark##1{}
 \def\subsectionmark##1{}}
\def\ps@titlepage{\let\@mkboth\@gobbletwo
 \def\@oddhead{\footnotesize\@journal\hfill}
 \def\@oddfoot{}
 \def\@evenhead{\footnotesize\@journal\hfill}
 \def\@evenfoot{}
 \def\sectionmark##1{}
 \def\subsectionmark##1{}}

\def\@pnumwidth{1.55em}
\def\@tocrmarg {2.55em}
\def\@dotsep{4.5}
\def\@undottedtocline#1#2#3#4#5{\ifnum #1>\c@tocdepth
 \else
  \vskip \z@ plus .2pt
  {\hangindent #2\relax
   \rightskip \@tocrmarg \parfillskip -\rightskip
   \parindent #2\relax \@afterindenttrue
   \interlinepenalty\@M \leavevmode
   \@tempdima #3\relax #4\nobreak \hfill \nobreak
   \hbox to\@pnumwidth{\hfil\rm \ }\par}\fi}
\def\tableofcontents{\@restonecolfalse
 \if@twocolumn\@restonecoltrue\onecolumn\fi
 \section*{CONTENTS} \@starttoc{toc}
 \if@restonecol\twocolumn\fi \par\vspace{12pt}}
\def\l@part#1#2{\addpenalty{-\@highpenalty}
 \addvspace{2.25em plus 1pt}
 \begingroup
  \parindent \z@ \rightskip \@pnumwidth
  \parfillskip -\@pnumwidth
  {\normalsize\rm
   \leavevmode \hspace*{3pc}
   #1\hfil \hbox to\@pnumwidth{\hss \ }}\par
   \nobreak \global\@nobreaktrue
   \everypar{\global\@nobreakfalse\everypar{}}\endgroup}
\def\l@section#1#2{\addpenalty{\@secpenalty}
 \@tempdima 1.5em
 \begingroup
  \parindent \z@ \rightskip \@pnumwidth
  \parfillskip -\@pnumwidth \rm \leavevmode
  \advance\leftskip\@tempdima \hskip -\leftskip
  #1\nobreak\hfil \nobreak\hbox to\@pnumwidth{\hss \ }\par
 \endgroup}
\def\l@subsection{\@undottedtocline{2}{1.5em}{2.3em}}
\def\l@subsubsection{\@undottedtocline{3}{3.8em}{3.2em}}
\def\l@paragraph{\@undottedtocline{4}{7.0em}{4.1em}}
\def\l@subparagraph{\@undottedtocline{5}{10em}{5em}}
\def\listoffigures{\@restonecolfalse
 \if@twocolumn\@restonecoltrue\onecolumn\fi
 \section*{LIST OF FIGURES\@mkboth{LIST OF FIGURES}{LIST OF FIGURES}}
 \@starttoc{lof} \if@restonecol\twocolumn\fi}
\def\l@figure{\@undottedtocline{1}{1.5em}{2.3em}}
\def\listoftables{\@restonecolfalse
 \if@twocolumn\@restonecoltrue\onecolumn\fi
 \section*{LIST OF TABLES\@mkboth{LIST OF TABLES}{LIST OF TABLES}}
 \@starttoc{lot} \if@restonecol\twocolumn\fi}
\let\l@table\l@figure

\def\thebibliography#1{\section*{REFERENCES}
 \addcontentsline{toc}{section}{REFERENCES}
 \list{}{\labelwidth\z@
         \leftmargin 1.5em
	 \itemsep \z@
	 \itemindent-\leftmargin}
 \small\raggedright
 \parindent\z@
 \parskip\z@ plus .1pt\relax
 \def\newblock{\hskip .11em plus .33em minus .07em}
 \sloppy\clubpenalty4000\widowpenalty4000
 \sfcode`\.=1000\relax
}

\def\@biblabel#1{\hspace*{\labelsep}[#1]}

\newif\if@restonecol
\def\theindex{\section*{INDEX}
 \addcontentsline{toc}{section}{INDEX}
 \footnotesize \parindent\z@ \parskip\z@ plus .1pt\relax
 \let\item\@idxitem}
\def\@idxitem{\par\hangindent 1em}

\def\endtheindex{\if@restonecol\onecolumn\else\clearpage\fi}

\def\footnoterule{\kern-3\p@ \hrule width 12pc height \z@ \kern 3\p@}

\def\@fnsymbol#1{\ifcase#1\or \mbox{$\star$}\or \dagger\or \ddagger\or
   \S \or \P \or \|\or **\or \dagger\dagger
   \or \ddagger\ddagger\or \S\S\or \P\P\or \|\|\else ***
   \fi\relax}

\long\def\@makefntext#1{\parindent 1em\noindent
  $^{\@thefnmark}$\hspace{4pt}#1}

\newcounter{table}
\def\thetable{\@arabic\c@table}
\def\fps@table{tbp}
\def\ftype@table{1}
\def\ext@table{lot}
\def\fnum@table{Table \thetable}
\def\table{\let\@makecaption=\SFB@maketablecaption\@float{table}}
\let\endtable\end@float
\@namedef{table*}{\let\@makecaption=\SFB@maketablecaption\@dblfloat{table}}
\@namedef{endtable*}{\end@dblfloat}

\newcounter{figure}
\def\thefigure{\@arabic\c@figure}
\def\fps@figure{tbp}
\def\ftype@figure{2}
\def\ext@figure{lof}
\def\fnum@figure{Figure \thefigure}
\def\figure{\let\@makecaption=\SFB@makefigurecaption\@float{figure}}
\let\endfigure\end@float
\@namedef{figure*}{\let\@makecaption=\SFB@makefigurecaption\@dblfloat{figure}}
\@namedef{endfigure*}{\end@dblfloat}

\long\def\SFB@makefigurecaption#1#2{\vskip 6pt
 \setbox\@tempboxa\hbox{\small{\bf #1.} #2}
 \ifdim \wd\@tempboxa >\hsize
  \small{\bf #1.} #2\par
 \else
  \hbox to\hsize{\hfil\box\@tempboxa\hfil}
 \fi
 \vskip 6pt
}
\long\def\SFB@maketablecaption#1#2{\vskip 6pt
 \setbox\@tempboxa\hbox{\small{\bf #1.} #2}
 \ifdim \wd\@tempboxa >\hsize
  \small{\bf #1.} #2\par
 \else
  \hbox to\hsize{\box\@tempboxa\hfill}
 \fi
 \vskip 6pt
}

\def\caption{\@ifstar{\SFB@caption\@captype}%
 {\refstepcounter\@captype \@dblarg{\@caption\@captype}}%
}
\long\def\SFB@caption#1#2{%\par
 \begingroup
  \@parboxrestore
  \normalsize
  \@makecaption{\csname fnum@#1\endcsname}{\ignorespaces #2}\par
 \endgroup}

\def\@cite#1#2{(#1\if@tempswa , #2\fi)}
\def\@biblabel#1{}

\newlength{\bibhang}
\def\@citex[#1]#2{\if@filesw\immediate\write\@auxout{\string\citation{#2}}\fi
  \def\@citea{}\@cite{\@for\@citeb:=#2\do
    {\@citea\def\@citea{; }\@ifundefined
       {b@\@citeb}{{\bf ?}\@warning
       {Citation `\@citeb' on page \thepage \space undefined}}%
{\csname b@\@citeb\endcsname}}}{#1}}
\let\@internalcite\cite
\def\cite{\def\citename##1{##1}\@internalcite}
\def\shortcite{\def\citename##1{}\@internalcite}

\def\[{\relax\ifmmode\@badmath\else\begin{trivlist}\item[]\leavevmode
  \hbox to\linewidth\bgroup$
  \displaystyle
  \hskip\mathindent\bgroup\fi}

\def\]{\relax\ifmmode \egroup $\hfil
       \egroup \end{trivlist}\else \@badmath \fi}

\def\equation{\refstepcounter{equation}\trivlist \item[]\leavevmode
  \hbox to\linewidth\bgroup $
  \displaystyle
\hskip\mathindent}

\def\endequation{$\hfil
           \displaywidth\linewidth\@eqnnum\egroup \endtrivlist}

\def\eqnarray{\stepcounter{equation}\let\@currentlabel=\theequation
\global\@eqnswtrue
\global\@eqcnt\z@\tabskip\mathindent\let\\=\@eqncr
\abovedisplayskip\topsep\ifvmode\advance\abovedisplayskip\partopsep\fi
\belowdisplayskip\abovedisplayskip
\belowdisplayshortskip\abovedisplayskip
\abovedisplayshortskip\abovedisplayskip
$$\halign
to \linewidth\bgroup\@eqnsel\hskip\@centering$\displaystyle\tabskip\z@
  {##}$&\global\@eqcnt\@ne \hskip 2\arraycolsep \hfil${##}$\hfil
  &\global\@eqcnt\tw@ \hskip 2\arraycolsep $\displaystyle{##}$\hfil
   \tabskip\@centering&\llap{##}\tabskip\z@\cr}

\def\endeqnarray{\@@eqncr\egroup
 \global\advance\c@equation\m@ne$$\global\@ignoretrue}

\newdimen\mathindent
\mathindent = \z@

\def\today{\number\day\ \ifcase\month\or
  January\or February\or March\or April\or May\or June\or
  July\or August\or September\or October\or November\or December
 \fi \ \number\year}

\ps@headings
\ifSFB@galley
 \ps@empty
\fi
\ifSFB@referee
\fi
\if@twocolumn
 \twocolumn
\else
 \onecolumn
\fi
\documentstyle{mn}

\def\eg{e.g.\/}
\def\etal{et~al.\/}
\def\g{$\gamma$}

\newbox\grsign \setbox\grsign=\hbox{$>$}
\newdimen\grdimen \grdimen=\ht\grsign
\newbox\laxbox \newbox\gaxbox
\setbox\gaxbox=\hbox{\raise.5ex\hbox{$>$}\llap
     {\lower.5ex\hbox{$\sim$}}}\ht1=\grdimen\dp1=0pt
\setbox\laxbox=\hbox{\raise.5ex\hbox{$<$}\llap
     {\lower.5ex\hbox{$\sim$}}}\ht2=\grdimen\dp2=0pt
\def\gax{\mathrel{\copy\gaxbox}}
\def\lax{\mathrel{\copy\laxbox}}
\def\simless{\lax}
\def\simgreat{\gax}

\def\c{(C)_{\rm min}}

\def\C{(C)_{\rm max}}

\def\Cbarsix{(\bar{C}^{64})_{\rm max}}

\def\Cbarten{(\bar{C}^{1024})_{\rm max}}

\def\Bmid{B_{\rm mid}}

\begin{document}

\title[Evidence for the Galactic origin of $\gamma$-ray bursts]
      {Evidence for the Galactic origin of $\gamma$-ray bursts}

\author
[J. M. Quashnock and D. Q. Lamb]
	{J. M. Quashnock and D. Q. Lamb \\
	Department of Astronomy and Astrophysics, University of Chicago,
	Chicago, IL 60637}

\date{Accepted 30 September 1993; received 20 September 1993; received
in original form 15 June 1993.}

\maketitle

\begin{abstract}

We investigate the angular distribution of the $\gamma$-ray bursts in
the publicly available BATSE catalogue, using the measures of burst
brightness $B$ and short time scale ($\simless$ 0.3 s) variability $V$
which we introduced earlier.  We show that the 54 type I ($\log V \le
-0.8$) bursts lying in the middle brightness range 490 counts $\le B
\le$ 1250 counts (corresponding to 1/3 of all type I bursts) exhibit a
Galactic dipole moment of $\langle \cos \theta \rangle = 0.204 \pm
0.079$ {\it and} a deviation of the Galactic quadrupole moment from 1/3
of $\langle \sin^2 b \rangle - 1/3 = -0.104 \pm 0.041$.  Using Monte
Carlo simulations which include the BATSE sky exposure map and taking
into account division of the type I bursts into three equal samples, we
find that the probability by chance of an isotropic distribution of 54
bursts exhibiting values of $\langle \cos \theta \rangle$ {\it and} the
negative of $\langle \sin^2 b \rangle - 1/3$ which equal or exceed the
observed values is $6.6 \times 10^{-5}$.  We conclude that $\gamma$-ray
bursts are Galactic in origin.

\end{abstract}


\section{Introduction}

Gamma-ray bursts continue to confound astrophysicists a quarter of a
century after their discovery (Klebesadel, Strong, and Olson 1973).
The challenge of deciphering the nature of the bursts is exacerbated by
the fact that one cannot predict when or from where on the sky the
bursts will come, and the fact that it has been impossible to find
quiescent counterparts of the bursts at radio, infrared, optical,
ultraviolet, X-ray, or $\gamma$-ray energies.  The lack of quiescent
counterparts puts a premium on garnering knowledge from the bursts
themselves.  The distribution of bursts on the sky constrains the
spatial distribution of burst sources, and may indicate an astronomical
distance scale.

Prior to the launch of the {\it Compton} Observatory, it was generally
believed that \g-ray bursts come from neutron stars in the Galactic
disk, located at distances $\lax$ 300 pc with luminosities $\sim
10^{37}$ ergs~s$^{-1}$ (see, \eg, Higdon and Lingenfelter 1990; Harding
1992).  The confirmation of a roll-over in the brightness distribution
of $\gamma$-ray bursts and the discovery that even faint \g-ray bursts
appear to be isotropically distributed on the sky (Meegan et al. 1992)
has intensified debate about whether $\gamma$-ray bursts are Galactic
or cosmological in origin.

The BATSE catalogue (Fishman \etal\ 1993) is the largest homogeneous
sample of \g-ray bursts ever assembled and, as such, constitutes a
unique resource for studying the angular distribution of bursts.  Here
we investigate the Galactic dipole and quadrupole moments of the
angular distribution of $\gamma$-ray bursts, using the measures of
burst brightness $B = \Cbarten$ and short time scale ($\simless$ 0.3 s)
variability $V = \Cbarsix / \Cbarten$, which we introduced earlier
(Lamb, Graziani, and Smith 1993).  The quantities $\Cbarsix$ and
$\Cbarten$ are the expected maximum number of counts in 64 ms and in
1024 ms, respectively.

\section{Analysis}

In earlier studies (Lamb, Graziani, and Smith 1993; Lamb and Graziani
1993a,b), we presented evidence for two distinct morphological classes
of $\gamma$-ray bursts (see also Kouveliotou \etal\ 1993).  Type I
bursts (comprising $\approx$ 80\% of the bursts) are smoother ($\log V
\le -0.8$) on short time scales ($\simless 0.3$ s), range from faint to
bright, are longer, and have softer spectra.  Type II bursts
(comprising $\approx$ 20\% of the bursts) are more variable ($\log V >
-0.8$) on short time scales ($\simless 0.3$ s), faint ($B <$ 1900
counts), shorter, and have harder spectra.

In a separate paper (Quashnock and Lamb 1993), we study the clustering
of $\gamma$-ray bursts.  We show that all bursts and the 201 bursts for
which $B$ and $V$ exist are significantly clustered on an angular scale
$\approx 5^\circ$.  This angular scale is smaller than the typical
(statistical plus systematic) error in burst locations of $6.8^\circ$,
suggesting multiple recurrences from individual sources.  Further, we
show that the bright and faint bursts are correlated with each other,
while the medium brightness bursts are not.  These results imply that
``classical'' $\gamma$-ray burst sources repeat on a time scale of
months, and that many faint type I and II bursts come from the sources
of bright type I bursts (Quashnock and Lamb 1993).

Motivated by this knowledge, we investigate the angular distribution of
$\gamma$-ray bursts as a function of burst brightness $B$.  Peak
brightness, whether measured by $\Cbarsix$ or $B [=\Cbarten]$, appears
to be a better ``standard candle'' for type I bursts, which are smooth
on short time scales ($\simless 0.3$ s), than it is for type II bursts,
which are variable on short time scales (Lamb and Graziani 1993b).  In
addition, type II bursts are short and faint, resulting in a relatively
large statistical error (which is inversely proportional to the square
root of the fluence of the burst) in their location on the sky.

We therefore focus our attention on the angular distribution of type I
bursts and use $B$ as our measure of burst brightness.  Fig.~1 shows
the locations on the sky of the 163 type I bursts for which $B$ exists
in the publicly available BATSE catalogue (Fishman \etal\ 1993).  We
divide these type I bursts into three brightness samples containing
equal numbers of bursts: 54 bursts which have $B < 490$ counts, 54
bursts which have 490 counts $\le B \le$ 1250 counts, and 55 bursts
which have $B >$ 1250 counts.  We then compute the mean values of the
Galactic dipole moment $\langle \cos \theta \rangle$ and quadrupole
moment $\langle \sin^2 b \rangle$ for each sample.

\section{Results}

Table 1 gives the results.  For an isotropic source distribution, we
expect $\langle \cos \theta \rangle = 0$ and $\langle \sin^2 b \rangle
- 1/3 = 0$, irrespective of brightness $B$; corrected for the BATSE sky
exposure map (Brock et al. 1992, Fishman et al. 1993), these become
$\langle \cos \theta \rangle = -0.0123$ and $\langle \sin^2 b \rangle -
1/3 = - 0.0047$.  We find that the angular distributions of the faint
and bright samples of type I bursts are consistent with isotropy,
whereas the 54 type I bursts in the middle brightness sample have large
Galactic dipole and quadrupole moments.  We evaluate the significance
of these moments using Monte Carlo simulations.  In each simulation, we
draw 54 bursts at random from the flux-integrated BATSE sky exposure
map, which is symmetric in Right Ascension and a function of
Declination (Brock et al. 1992, Fishman et al. 1993).  We find that the
probability by chance of an isotropic distribution of 54 bursts
exhibiting values of $\langle \cos \theta \rangle$ {\it and} the
negative of $\langle \sin^2 b \rangle - 1/3$ that equal or exceed the
observed values is $2.2 \times 10^{-5}$ (see Table 2).  Multiplying
this result by a factor of 3 in order to take into account having
divided the type I bursts into three equal samples, we obtain a
significance of $6.6 \times 10^{-5}$.

Having established significant evidence for a concentration of middle
brightness type I bursts toward the Galactic center and in the Galactic
plane, we construct scatter plots and calculate running averages of the
Galactic dipole and quadrupole moments in order to explore further
their behavior as a function of $B$.  Fig.~2 shows the distribution of
bursts in the ($B,\cos \theta$)-diagram and in the (B,$\sin
b$)-diagram, and the running averages of $\langle \cos \theta \rangle$
and $\langle \sin^2 b \rangle - 1/3$ as a function of burst brightness
$B$ using a brightness window $\Delta \log B = 0.4$.

The scatter plot of $\cos \theta$ and the running average of $\langle
\cos \theta \rangle$ show that type I bursts in the brightness range
500 counts $\simless B \simless$ 2000 counts are concentrated toward
the Galactic center, and hint that those in the brightness range 3000
counts $\simless B \simless$ 6000 counts lie preferentially toward the
Galactic anticenter.  The scatter plot of $\sin b$ and the running
average of $\langle \sin^2 b \rangle - 1/3$ show that type I bursts in
the brightness range 500 counts $\simless B \simless$ 1300 counts are
concentrated in the Galactic plane, and hint that those in the
brightness range 2000 counts $\simless B \simless$ 4000 counts lie
preferentially in the Galactic plane while those in the brightness
range 4000 counts $\simless B \simless$ 9000 counts lie preferentially
at high Galactic latitudes.  In addition, the ($B, \sin b$)-diagram
exhibits a general symmetry between positive and negative $\sin b$.

As a result of this complex behavior, the Galactic dipole and
quadrupole moments of the full sample of type I bursts do not differ
significantly from the values expected for isotropy (see Table 1).  We
note further that previous experiments were sensitive only to bright
bursts, and therefore did not detect many of the middle brightness type
I bursts in which we find a concentration toward the Galactic center
and in the Galactic plane.

The running average of the dipole moment peaks at $\langle \cos \theta
\rangle = 0.230 \pm 0.078$ when $\Bmid = 737$, corresponding to 55
bursts in the medium brightness range 465 counts $< B <$ 1169 counts.
The running average of the deviation of the quadrupole moment from 1/3
is $\langle \sin^2 b \rangle - 1/3 = -0.119 \pm 0.040$ for the same
$\Bmid$.  These values of the Galactic dipole and quadrupole moments
{\it each} represent 2.96 $\sigma$ deviations from the values expected
for isotropy.  Fig.~3 shows the distribution on the sky of these 55
bursts.  Fifty-one of the 54 bursts in the middle brightness range 490
counts $< B <$ 1250 counts are common to the 55 in the medium
brightness range 465 counts $< B <$ 1169 counts.

Considering {\it only} the 55 bursts in the medium brightness range 465
counts $< B <$ 1169 counts and again using Monte Carlo simulations
which take into account the BATSE sky exposure map, we calculate that
the probability by chance of an isotropic distribution of 55 bursts
exhibiting values of $\langle \cos \theta \rangle$ {\it and} the
negative of $\langle \sin^2 b \rangle - 1/3$ that jointly equal or
exceed the above values is $1.9 \times 10^{-6}$.

This significance does not take into account the large number of
independent trials introduced by performing a running average over the
brightness range $2 \le \log B \le 4.5$.  In order to do so, we
consider all the bursts and use more elaborate Monte Carlo
simulations.  In each simulation, we draw 163 bursts at random from the
flux-integrated BATSE sky exposure map, assign a brightness $B$ to each
from the set of 163 actual brightnesses, and calculate burst-by-burst
running averages of the Galactic dipole and quadrupole moments using a
brightness window $\Delta \log B = 0.4$, as we did for the BATSE data.
We then determine the fraction of the simulations for which a window
{\it anywhere} in the simulation exhibits values divided by their
uncertainties of $\langle \cos \theta \rangle$ {\it and} the negative
of $\langle \sin^2 b \rangle - 1/3$ that equal or exceed the observed
values.  We obtain a significance of $1.1 \times 10^{-4}$.

The running average of $\langle \sin^2 b \rangle - 1/3$ shows that the
fainter bursts in this medium brightness range are more strongly
concentrated in the Galactic plane.  Indeed, the nine faintest bursts
in this brightness range (denoted by open circles in Figs. 2 and 3) are
characterized by $\langle \sin^2 b \rangle - 1/3 = -0.280 \pm 0.099$
(representing a 2.8 $\sigma$ deviation from the value expected for
isotropy for these bursts alone).

We have also examined the angular distribution of type II bursts.
While there are hints that the behavior of the angular distribution of
type II bursts as a function of brightness may be similar to that of
type I bursts, the evidence is not statistically significant.

\section{Discussion}

We find that the 54 type I bursts in the middle brightness range 490
counts $< B <$ 1250 counts (which constitute 1/3 of type I bursts) are
strongly concentrated toward the Galactic center and in the Galactic
plane.  We conclude that these type I bursts, and thus all type I
bursts, are Galactic in origin.  As mentioned earlier, in a separate
paper we report evidence that many faint type I and type II bursts come
from the sources of bright type I bursts (Quashnock and Lamb 1993).  We
therefore conclude that all $\gamma$-ray bursts are Galactic in
origin.

Fig.~2 hints that the faint and bright type I bursts lie preferentially
toward the anti-center and at high Galactic latitudes around us.
Figs.~2 and 3 show that type I bursts in the medium brightness range
465 counts $< B <$ 1169 are strongly concentrated toward the Galactic
center and in the Galactic plane, with the faintest of the medium
brightness bursts concentrated very strongly in the Galactic plane.
This distribution suggests that most type I burst sources are
associated with the Galactic disk.

We infer that BATSE sees only a fraction of the type I burst sources in
the Galaxy.  Otherwise, we would expect that the concentration of
bursts toward the Galactic center and in the Galactic plane would be
very strong, which is not the case (see Fig.  3).  Further, we would
expect, qualitatively, that the cumulative brightness distribution of
type I bursts would have a slope $\propto -1$ throughout most of its
range, whereas it is flatter at the faint end and steeper at the bright
end (Lamb and Graziani 1993b).

We therefore conjecture that BATSE is seeing type I burst sources out
to $\sim 1 - 2$ kpc, and that the behavior of the angular distribution
of these bursts as a function of burst brightness $B$ reflects the
structure of the nearby spiral arms of the Galaxy.  In this picture,
the bright type I bursts come primarily from the vicinity of the Orion
arm, which lies around us and toward the Galactic anticenter (at
distances of up to $\approx$ 1 kpc).  The medium brightness type I
bursts come primarily from the vicinity of the Sagittarius arm, which
lies toward the Galactic center (at a distance $\approx$ 1 - 1.5 kpc).
Some of the faint type I bursts may come from the vicinity of the
Perseus arm, which lies toward the Galactic anticenter (at distances
$\simgreat$ 2 kpc).

Figs.~2 and 3 hint that the faint type I bursts lie preferentially
toward the Galactic anticenter but not in the Galactic plane.  Although
the uncertainties in the locations of many of the faint type I bursts
are $10^\circ - 15^\circ$, they are not sufficient to mask the strong
concentration in the Galactic plane which would be expected if most of
the faint type I bursts came from the Perseus arm.  One possibility is
that the faint bursts are part of a roughly isotropic source population
upon which the angular distribution of the Galactic disk is
superimposed (Smith and Lamb 1993); however, the existence of such a
roughly isotropic source population is contradicted by the lack of
medium brightness type I bursts at high Galactic latitudes.  An
alternative possibility is that the faint type I bursts come from
nearby sources in the vicinity of the Orion arm.  In a separate paper,
we present evidence that this is indeed the case, and that many of the
faint type I and II bursts come from the sources of bright type I
bursts (Quashnock and Lamb 1993).

Such a result implies that the luminosity function of $\gamma$-ray
burst sources is broad.  Nevertheless, the strong variations in the
Galactic dipole and quadrupole moments of type I bursts as a function
of $B$ (which occur within $\Delta \log B \approx 0.6$) imply that the
(peak) brightness of many bright type I bursts is a surprisingly good
``standard candle'' and therefore provides a good estimate of the
distance to the burst source.  Otherwise the variations in the angular
distribution would be washed out by the scatter in the intrinsic
luminosity of the bursts.  Indeed, the ``notch'' at $\C / \c = 6$ in
the cumulative $\C / \c$ distribution of type I bursts (Lamb and
Graziani 1993b) may reflect the difference between the typical distance
to sources in the Orion spiral arm and that to sources in the
Sagittarius spiral arm.  These results support the inference we drew
from our earlier study of the cumulative brightness distribution of
$\gamma$-ray bursts that the intrinsic luminosity of the bursts may
``saturate'' at some value (Lamb and Graziani 1993b; see also Mao,
Narayan, and Piran 1993).  Within the framework of our conjecture that
the $\gamma$-ray bursts seen by BATSE come from the vicinity of the
nearby spiral arms of the Galaxy, this value is roughly the Eddington
luminosity of a solar mass object; namely, $L_E \approx 2 \times
10^{38}$ erg s$^{-1}$.

Neutron stars are an attractive possibility for the sources of
$\gamma$-ray bursts (see, e.g., Lamb 1983).   We conjecture that the
bursts come from young neutron stars which have not yet left the
vicinity of the spiral arms of the Galaxy or from old neutron stars
which become active burst sources as they pass through the spiral arms
of the Galaxy, possibly by accreting matter from clouds.

Most neutron stars are thought to have velocities $v \approx 100 - 200$
km s$^{-1}$.  If many of the medium brightness type I bursts come from
the vicinity the Sagittarius spiral arm, as we conjecture, the Galactic
latitudes of these bursts imply that burst sources lie as much as 0.2 -
0.5 kpc above and below the Galactic plane (the uncertainties in the
locations of the bursts are not sufficient to account for the spread in
Galactic latitude, if the sources were to lie entirely in the plane).
This is comparable to the scale height $h \sim 0.2 - 0.5$ kpc expected
for neutron stars (Hartmann, Epstein, and Woosley 1990; Paczynski 1990)
and suggests, given neutron star velocities, that burst activity,
whether in young or old neutron stars, persists for as much as $\sim
10^6$ years.  Given such velocities, neutron stars leave the vicinity
of the Galactic spiral arms in a time $t_{\rm escape} \simless 10^7$
yrs.  Therefore burst activity, whether in young or old neutron stars,
cannot persist for longer than this without washing out the structure
in the angular and brightness distributions arising from the nearby
spiral arms of the Galaxy.

After this study was largely complete, we learned that Atteia \&
Dezalay (1993) have analyzed the behavior of $\langle V/V_{\rm max}
\rangle$ as a function of detector threshold by truncating the
cumulative $\C / \c$ distribution for bursts in the publicly available
BATSE catalogue (Fishman et al. 1993) at successively larger values of
$\C / \c$.  They find that $\langle V/V_{\rm max} \rangle$ is not a
monotonic function of detector threshold, but rather exhibits a sharp
dip when the threshold is set such that only the 50 brightest bursts
remain.  This dip in $\langle V/V_{\rm max} \rangle$ corresponds to the
``notch'' at $\C / \c = 6$ in the cumulative $\C / \c$ distribution of
type I bursts (Lamb and Graziani 1993b).  Examining the angular
distribution of the 24 bursts with $\C / \c$ values just above the dip,
they find no evidence for a Galactic dipole moment and modest (2.7
$\sigma$) evidence for a deviation of the Galactic quadrupole moment
from the value of 1/3 expected for isotropy.  This result corresponds
to the dip we find at $\Bmid \approx 3000$ in the running average of
$\langle \sin^2 b \rangle - 1/3$ (see the lower right panel in Fig.~2).

\subsection*{Acknowledgments}

We gratefully acknowledge the contributions of the scientists who
designed, built, and flew BATSE on the {\it Compton} Observatory, and
whose efforts made possible the work reported here.  We thank Paolo
Coppi for stimulating discussions about the spatial distribution of
$\gamma$-ray bursts, and Carlo Graziani and Tom Loredo for informative
discussions about statistical methodology.  Finally, we thank the
anonymous referee for valuable comments.  This research was supported
in part by NASA grants NAGW-830, NAGW-1284, and NASW-4690.

\vfill\eject

\null
\vfill\eject

\begin{figure}
\centering
\caption{
Distribution on the sky of 163 type I bursts for which the brightness
$B$ exists in the publicly available BATSE catalogue (Fishman
\etal\ 1993), in Galactic coordinates (the Galactic Center lies at the
center of the map).
}
\vskip 2cm
\end{figure}

\begin{figure}
\centering
\caption{
(upper left-hand panel)
Distribution of 163 type I bursts in the ($B,\cos \theta$)-diagram.
Note the concentration of bursts toward the Galactic center in the
brightness range 500 counts $\simless B \simless$ 2000 counts and the
hint that bursts in the brightness range 3000 counts $\simless B
\simless$ 6000 counts lie preferentially toward the Galactic
anticenter.  The bursts denoted by open circles are the faintest nine
bursts of 55 in the brightness range 465 counts $< B <$ 1169 counts and
have $\langle \sin^2 b \rangle -1/3 = -0.280 \pm 0.099$ (see Fig.~3).
(lower left-hand panel)
Running average of the Galactic dipole moment $\langle \cos \theta
\rangle$ as a function of $B$ using a brightness window $\Delta \log B =
0.4$.  The Galactic dipole moment (uncorrected for the BATSE sky
exposure map) peaks at $\langle \cos \theta \rangle = 0.230 \pm 0.078$
when $\Bmid = 737$.
(upper right-hand panel)
Distribution of 163 type I bursts in the ($B,\sin b$)-diagram.  Note
the concentration of bursts in the Galactic plane in the brightness
range 500 counts $\simless B \simless$ 1300 counts, and the hints that
those in the brightness range 2000 counts $\simless B \simless$ 4000
counts lie preferentially in the Galactic plane while those in the
brightness range 4000 counts $\simless B \simless$ 9000 counts lie
preferentially at high Galactic latitudes.  The nine bursts denoted by
open circles are the same as those in the upper left-hand panel.
(lower right-hand panel)
Running average of the deviation of the Galactic quadrupole moment from
1/3, $\langle \sin^2 b \rangle - 1/3$, as a function of $B$ using a
brightness window $\Delta \log B = 0.4$.  The deviation of the Galactic
quadrupole moment from 1/3 (uncorrected for the BATSE sky exposure map)
is $\langle \sin^2 b \rangle - 1/3 = -0.119 \pm 0.040$ when $\Bmid =
737$.
}
\end{figure}

\begin{figure}
\centering
\caption{
Distribution on the sky of the 55 type I bursts in the medium
brightness range $465 < B < 1169$, in Galactic coordinates (the
Galactic Center lies at the center of the map).  The bursts denoted by
open circles are the nine faintest bursts in this brightness range;
they are characterized by $\langle \sin^2 b \rangle - 1/3 = -0.280 \pm
0.099$ (see also Fig.~2).
}
\vskip 4cm
\end{figure}

\null
\vfill\eject

\onecolumn
\begin{table}
\centering
\caption{Galactic dipole and quadrupole moments for different
brightness samples of type I bursts.}
\medskip
\begin{tabular}{ccccc}
& \multicolumn{4}{c}{Brightness range (counts)}\\
Moment& $\quad B<490$ & $490<B<1250$ & $B>1250$ &
\quad Total\\
&(54 bursts)&(54 bursts)&(55 bursts)&(163 bursts)\\
\hline
\\
$\langle\cos\theta\rangle$&$-0.041\pm 0.079$&$+0.204\pm 0.079$&$-0.059\pm
0.078$
&$+0.034\pm 0.045$\\
\\
$\langle\sin^2b\rangle - {1\over 3}$&$-0.010\pm 0.041$&$-0.104\pm 0.041$&
$+0.008\pm 0.040$&$-0.035\pm 0.023$\\
\\
\hline
\end{tabular}
\end{table}

\begin{table}
\centering
\caption{Galactic dipole and quadrupole moments for the 54 type~I
bursts in the middle brightness range 490 counts $<B<$ 1250 counts, and
the corresponding Q-values, corrected for the BATSE sky exposure map.}
\medskip
\begin{tabular}{ccccc}
$\langle\cos\theta\rangle$&$\langle\sin^2b\rangle-{1\over 3}$&$Q_{\cos}$
&$Q_{\sin}$&$Q_{\rm both}$\\
\hline
\\
$+0.204\pm 0.079$&$-0.104\pm 0.041$&$2.4\times 10^{-3}$&$5.7 \times
10^{-3}$&$2.2\times 10^{-5}$\\
\\
\hline
\end{tabular}
\end{table}


\begin{thebibliography}{}

\bibitem[Atteia and Dezalay \ 1993]{F93}
Atteia, J.-L. \& Dezalay, J.-P. 1993, A\&A, 274, L1.

\bibitem[Brock et al. \ 1992]{B92}
Brock, M., N., Meegan, C. A., Fishman, G. J., Wilson, R. B., Paciesas,
W. S., \& Pendleton, G. N.  1992, in {\it Gamma-Ray Bursts}, eds. W.
S.  Paciesas \& G. J. Fishman (New York: AIP), p. 399.

\bibitem[Fishman et al. \ 1993]{F93}
Fishman, G. J., et al. 1993, ApJSup, submitted.

\bibitem[Harding 1991]{H91}
Harding, A. 1992, Phys. Rep. 206, 327.

\bibitem[Hartmann et al. / 1990]{H90}
Hartmann, D., Epstein, R. I., \& Woosley, S. E. 1990, ApJ, 348, 625.

\bibitem[Higdon \& Lingenfelter 1990]{HL90}
Higdon, J. C., \& Lingenfelter, R. E. 1990, Ann. Rev. Astron. Ap., 28, 401.

\bibitem[Klebesadel et al. \ 1973]{K73}
Klebesadel, R. W., Strong, I. B., \& Olson, R. A. 1973, ApJ, 182, L85.

\bibitem[Kouveliotou et al. \ 1993]{K93}
Kouveliotou, C., \etal\ 1993, ApJ, 413, L101.

\bibitem[Lamb 1983]{L83}
Lamb, D. Q. 1983, Ann. N.Y. Acad. Sci., 422, 237.

\bibitem[Lamb and Graziani \ 1993a]{LG93a}
Lamb, D. Q., \& Graziani, C. 1993a, ApJ, in press.

\bibitem[Lamb and Graziani \ 1993b]{LG93b}
Lamb, D. Q., \& Graziani, C. 1993b, ApJ, in press.

\bibitem[Lamb et al. \ 1993]{LG93}
Lamb, D. Q., Graziani, C., \& Smith, I. A. 1993, ApJ, 413, L11.

\bibitem[Mao et al. \ 1993]{M93}
Mao, S., Narayan, R., \& Piran, T.  1993, ApJ, submitted.

\bibitem[Meegan et al. \ 1992]{M92}
Meegan, C. A., Fishman, G. J., Wilson, R. B., Paciesas, W. S.,
Pendleton, G. N., Horack, J. M., Brock, M. N., \& Kouveliotou, C.
1992, Nature, 355, 143.

\bibitem[Paczynski 1990]{P90}
Paczynski, B. 1990, ApJ, 348, 485.

\bibitem[Press et al. \ 1986]{P86}
Press, W., Flannery, B. P., Teukolsky, S. A., \& Vetterling, W. T.
1986, {\it Numerical Recipes} Cambridge:  Cambridge University Press),
p. 472.

\bibitem[Quashnock and Lamb 1993]{QL93}
Quashnock, J. M., \& Lamb, D. Q.  1993, MNRAS, submitted.

\bibitem[Smith and Lamb \ 1993]{SL93}
Smith, I. A., \& Lamb, D. Q. 1993, ApJ, 410, L23.


\end{thebibliography}
\end{document}